\documentclass[12pt]{article}

\newcommand{\vp}{\vec {\it p}}
\newcommand{\be}{\begin{equation}}
\newcommand{\ee}{\end{equation}}
\newcommand{\bmt}{\begin{array}}
\newcommand{\emt}{\end{array}}
\newcommand{\muA}{\mu}
\newcommand{\muB}{\nu}
\newcommand{\muC}{\lambda}
\newcommand{\muD}{\rho}
\newcommand{\seqi}{{\muA}{\muB}{\muC}{\muD}}
\newcommand{\seqv}{{p_1},{p_2},{p_3},{p_4}}

\newcommand{\thm}{\tanh{\left(\frac{m}{2T}\right)}}
\newcommand{\thmN}[1]{\tanh^{#1}{\left(\frac{m}{2T}\right)}}
\newcommand{\ep}[3]{\epsilon^{#1 #2 #3}}
\newcommand{\intn}[1]{\int{\rm d}^{#1} x}
\newcommand{\beN}[1]{\be\label{#1}}

\begin{document}

\title{Parity-breaking electromagnetic interactions in thermal 
${\rm QED_3}$}
\author{F. T. Brandt$^\dagger$, Ashok Das$^\ddagger$ and J. Frenkel$^\dagger$ 
\\ \\
$^\dagger$Instituto de F\'{\i}sica,
Universidade de S\~ao Paulo\\
S\~ao Paulo, SP 05315-970, BRAZIL\\
$^\ddagger$Department of Physics and Astronomy,
University of Rochester\\
Rochester, NY 14627-0171, USA}
\maketitle

\bigskip
\noindent

\begin{abstract}
We examine the parity violating terms generated by the {\it box
diagram} in ${\rm QED_{3}}$
at finite temperature. These lead to both {\it extensive} as well as {\it
non-extensive}
effective actions, which have very distinct behavior in the 
long wavelength 
and static limits. We discuss a {\it large gauge Ward identity}
for the leading terms in the static limit, whose solution coincides
with the effective action proposed earlier.
\end{abstract}
\vfill\eject
Three-dimensional gauge theories coupled to matter are of interest in
the high-temperature domain of four-dimensional field theories, as well
as in the description of {\it planar phenomena} in condensed matter
physics \cite{dunne:1988lh}. An important feature of these theories is
the parity anomaly which manifests in the generation of a Chern-Simons
term in the effective  gauge field action, when the fermions are
integrated out. The simplest parity-breaking Chern-Simons term appears
in  ${\rm QED_3}$ at the level of the self-energy  and has the form
\beN{eq1}
I_{CS}=C(T)\,\intn{3}\,\ep{\mu}{\nu}{\lambda}\,A_\lambda F_{\mu\nu},
\ee
where $C(T)$ is a temperature-dependent coefficient
\cite{babu:1987rs}. It is  well known
that such terms, in a non-Abelian gauge theory, are not invariant
under {\it large gauge} transformations,
which are associated with nonzero topological winding numbers
\cite{deser:1982}.
This happens because, at arbitrary temperatures, $C(T)$ cannot be
chosen to  take discrete
values as required by the {\it large gauge} invariance of the
theory. Note that since, at finite temperature, the
time direction becomes compact, nontrivial gauge transformations can
arise even in an Abelian theory.

Recently much progress has been made in the
understanding of {\it large gauge} invariance at finite
temperature. In  particular, a mechanism was
presented in the context of an exactly soluble $0+1$ dimensional
model \cite{DDL}, where {\it large gauge} invariance was noted to be
restored in 
the full effective action. In other words, it was noted, both
perturbatively \cite{DD} as well as in the exact effective action
\cite{DDL} that although the Chern-Simons term violates {\it large
gauge} invariance, at finite temperature, there arise other terms in
the effective action restoring the invariance. 
An interesting feature of this mechanism, in the $0+1$ dimensional
finite  temperature
actions, is the presence of terms which are not simply space-time
integrals of a density  (the {\it non-extensive terms}).

These ideas have subsequently been extended to the $2+1$ dimensional
${\rm QED}$ theory, in the special gauge background $A_0=A_0(t)$ and 
$A_i=A_i(\vec x)$, where several aspects of the effective action have
been studied \cite{fosco:1997ei}. However, there is some reason to
believe that the actions obtained in such backgrounds may not
represent the complete action. 
This is because such a choice of gauge backgrounds may be too
restrictive. Furthermore, while in
such actions only non-extensive terms are present, it is known that in
the $1+1$ dimensional theory, the finite temperature effective action is
extensive (but non-local) \cite{DA}. Hence, one would expect such features to
appear also in the complete effective action of ${\rm QED_{3}}$, in a general
gauge background. Unfortunately, in this case, it is not possible to
determine the complete effective action in a closed form (unlike the
$0+1$ dimensional case). As a result, one has to resort to
perturbation theory \cite{DD}.

The main purpose of this work is to take a first step in this
direction, by analyzing, in perturbation theory, the behavior of the
parity-breaking part of the thermal electromagnetic 
{\it four-point function}. We first derive the analytic expression
for the corresponding effective action at zero temperature, which has a
Lorentz and gauge invariant form, involving only the electromagnetic field
tensor. However, at $T\neq 0$, the  four-point function is not
Lorentz invariant due to the presence of the heat bath, and
furthermore, one has to face the issue of non-analyticity of the
thermal amplitudes (which is not present in $0+1$ dimensions). The
calculation  of the four-point function, involves
a large number of terms in the intermediate steps and is extremely
difficult to carry out
in general. For this reason, we have studied the finite temperature
behavior of the box diagram only in two special, but important
limits, namely, in the {\it long wavelength}  and {\it static} limits. 
In the long wavelength limit, we find that the thermal contributions 
give rise to an extensive action which is manifestly invariant under
large  gauge
transformation. These terms have a leading
behavior proportional to $1/T$ at high temperature. In contrast, the
leading contributions, in the static limit, correspond to a non-extensive
action, which is in agreement with the form proposed  earlier in the
literature \cite{fosco:1997ei} and have a leading high
temperature  behavior proportional to $1/T^3$. 
There are, in addition extensive terms, which are suppressed  
at high temperature by extra powers of $1/T$.
Large gauge invariance is an important issue in the static limit. 
In this case, we can write for
the leading parity-breaking terms a non linear Ward identity, which
reflects the invariance of the full theory under {\it large gauge}
transformations. The solution of such a Ward identity coincides with
the non-extensive action proposed earlier in the specific static gauge
background. We describe next only the results of our analysis, leaving the
details  for a separate publication \cite{BDF}.

\begin{figure}[t]
\setlength{\unitlength}{0.007in}%
\begingroup\makeatletter\ifx\SetFigFont\undefined
\def\x#1#2#3#4#5#6#7\relax{\def\x{#1#2#3#4#5#6}}%
\expandafter\x\fmtname xxxxxx\relax \def\y{splain}%
\ifx\x\y   
\gdef\SetFigFont#1#2#3{%
  \ifnum #1<17\tiny\else \ifnum #1<20\small\else
  \ifnum #1<24\normalsize\else \ifnum #1<29\large\else
  \ifnum #1<34\Large\else \ifnum #1<41\LARGE\else
     \huge\fi\fi\fi\fi\fi\fi
  \csname #3\endcsname}%
\else
\gdef\SetFigFont#1#2#3{\begingroup
  \count@#1\relax \ifnum 25<\count@\count@25\fi
  \def\x{\endgroup\@setsize\SetFigFont{#2pt}}%
  \expandafter\x
    \csname \romannumeral\the\count@ pt\expandafter\endcsname
    \csname @\romannumeral\the\count@ pt\endcsname

  \csname #3\endcsname}%
\fi
\fi\endgroup
\begin{picture}(760,232)(60,290)
%
%
\thicklines
\multiput( 80,480)(3.2,-3.2){13}{\makebox(0.7,1.1){\SetFigFont{10}{12}{rm}.}}
\multiput(360,480)(3.2,-3.2){13}{\makebox(0.7,1.1){\SetFigFont{10}{12}{rm}.}}
\multiput(640,480)(3.2,-3.2){13}{\makebox(0.7,1.1){\SetFigFont{10}{12}{rm}.}}
\multiput(240,480)(-3.2,-3.2){13}{\makebox(0.7,1.1){\SetFigFont{10}{12}{rm}.}}
\multiput(520,480)(-3.2,-3.2){13}{\makebox(0.7,1.1){\SetFigFont{10}{12}{rm}.}}
\multiput(800,480)(-3.2,-3.2){13}{\makebox(0.7,1.1){\SetFigFont{10}{12}{rm}.}}
\multiput( 80,320)(3.2,3.2){13}{\makebox(0.7,1.1){\SetFigFont{10}{12}{rm}.}}
\multiput(360,320)(3.2,3.2){13}{\makebox(0.7,1.1){\SetFigFont{10}{12}{rm}.}}
\multiput(640,320)(3.2,3.2){13}{\makebox(0.7,1.1){\SetFigFont{10}{12}{rm}.}}
\multiput(240,320)(-3.2,3.2){13}{\makebox(0.7,1.1){\SetFigFont{10}{12}{rm}.}}
\multiput(520,320)(-3.2,3.2){13}{\makebox(0.7,1.1){\SetFigFont{10}{12}{rm}.}}
\multiput(800,320)(-3.2,3.2){13}{\makebox(0.7,1.1){\SetFigFont{10}{12}{rm}.}}
\put(120,360){\vector( 1, 0){ 45}}
\put(165,360){\line( 1, 0){ 35}}
\put(200,360){\vector( 0, 1){ 45}}
\put(200,405){\line( 0, 1){ 35}}
\put(200,440){\vector(-1, 0){ 45}}
\put(155,440){\line(-1, 0){ 35}}
\put(120,440){\vector( 0,-1){ 45}}
\put(400,360){\vector( 1, 0){ 45}}
\put(445,360){\line( 1, 0){ 35}}
\put(480,360){\vector( 0, 1){ 45}}
\put(480,405){\line( 0, 1){ 35}}
\put(480,440){\vector(-1, 0){ 45}}
\put(435,440){\line(-1, 0){ 35}}
\put(400,440){\vector( 0,-1){ 45}}
\put(680,360){\vector( 1, 0){ 45}}
\put(725,360){\line( 1, 0){ 35}}
\put(760,360){\vector( 0, 1){ 45}}
\put(760,405){\line( 0, 1){ 35}}
\put(760,440){\vector(-1, 0){ 45}}
\put(715,440){\line(-1, 0){ 35}}
\put(680,440){\vector( 0,-1){ 45}}
\put(120,360){\line( 0, 1){ 35}}
\put(400,360){\line( 0, 1){ 35}}
\put(680,360){\line( 0, 1){ 35}}
\put( 60,500){\makebox(0,0)[lb]{\smash{\SetFigFont{12}{14.4}{it}p}}}
\put( 80,500){\makebox(0,0)[lb]{\smash{\SetFigFont{12}{14.4}{rm},}}}
\put(220,500){\makebox(0,0)[lb]{\smash{\SetFigFont{12}{14.4}{it}p}}}
\put(240,500){\makebox(0,0)[lb]{\smash{\SetFigFont{12}{14.4}{rm},}}}
\put(340,500){\makebox(0,0)[lb]{\smash{\SetFigFont{12}{14.4}{it}p}}}
\put(360,500){\makebox(0,0)[lb]{\smash{\SetFigFont{12}{14.4}{rm},}}}
\put(500,500){\makebox(0,0)[lb]{\smash{\SetFigFont{12}{14.4}{it}p}}}
\put(520,500){\makebox(0,0)[lb]{\smash{\SetFigFont{12}{14.4}{rm},}}}
\put(620,500){\makebox(0,0)[lb]{\smash{\SetFigFont{12}{14.4}{it}p}}}
\put(640,500){\makebox(0,0)[lb]{\smash{\SetFigFont{12}{14.4}{rm},}}}
\put(780,500){\makebox(0,0)[lb]{\smash{\SetFigFont{12}{14.4}{it}p}}}
\put(800,500){\makebox(0,0)[lb]{\smash{\SetFigFont{12}{14.4}{rm},}}}
\put(100,500){\makebox(0,0)[lb]{\smash{\SetFigFont{12}{14.4}{rm}${\muD}$}}}
\put(260,500){\makebox(0,0)[lb]{\smash{\SetFigFont{12}{14.4}{rm}${\muC}$}}}
\put(380,500){\makebox(0,0)[lb]{\smash{\SetFigFont{12}{14.4}{rm}${\muD}$}}}
\put(540,500){\makebox(0,0)[lb]{\smash{\SetFigFont{12}{14.4}{rm}${\muB}$}}}
\put(660,500){\makebox(0,0)[lb]{\smash{\SetFigFont{12}{14.4}{rm}${\muC}$}}}
\put(820,500){\makebox(0,0)[lb]{\smash{\SetFigFont{12}{14.4}{rm}${\muD}$}}}
\put( 60,300){\makebox(0,0)[lb]{\smash{\SetFigFont{12}{14.4}{it}p}}}
\put( 80,300){\makebox(0,0)[lb]{\smash{\SetFigFont{12}{14.4}{rm},}}}
\put(100,300){\makebox(0,0)[lb]{\smash{\SetFigFont{12}{14.4}{rm}${\muA}$}}}
\put(340,300){\makebox(0,0)[lb]{\smash{\SetFigFont{12}{14.4}{it}p}}}
\put(360,300){\makebox(0,0)[lb]{\smash{\SetFigFont{12}{14.4}{rm},}}}
\put(380,300){\makebox(0,0)[lb]{\smash{\SetFigFont{12}{14.4}{rm}${\muA}$}}}
\put(620,300){\makebox(0,0)[lb]{\smash{\SetFigFont{12}{14.4}{it}p}}}
\put(640,300){\makebox(0,0)[lb]{\smash{\SetFigFont{12}{14.4}{rm},}}}
\put(660,300){\makebox(0,0)[lb]{\smash{\SetFigFont{12}{14.4}{rm}${\muA}$}}}
\put(780,300){\makebox(0,0)[lb]{\smash{\SetFigFont{12}{14.4}{it}p}}}
\put(800,300){\makebox(0,0)[lb]{\smash{\SetFigFont{12}{14.4}{rm},}}}
\put(820,300){\makebox(0,0)[lb]{\smash{\SetFigFont{12}{14.4}{rm}${\muB}$}}}
\put( 70,490){\makebox(0,0)[lb]{\smash{\SetFigFont{8}{8.0}{rm}4}}}
\put(230,490){\makebox(0,0)[lb]{\smash{\SetFigFont{8}{8.0}{rm}3}}}
\put(350,490){\makebox(0,0)[lb]{\smash{\SetFigFont{8}{8.0}{rm}4}}}
\put(510,490){\makebox(0,0)[lb]{\smash{\SetFigFont{8}{8.0}{rm}2}}}
\put(630,490){\makebox(0,0)[lb]{\smash{\SetFigFont{8}{8.0}{rm}3}}}
\put(790,490){\makebox(0,0)[lb]{\smash{\SetFigFont{8}{8.0}{rm}4}}}
\put( 70,290){\makebox(0,0)[lb]{\smash{\SetFigFont{8}{8.0}{rm}1}}}
\put(350,290){\makebox(0,0)[lb]{\smash{\SetFigFont{8}{8.0}{rm}1}}}
\put(630,290){\makebox(0,0)[lb]{\smash{\SetFigFont{8}{8.0}{rm}1}}}
\put(790,290){\makebox(0,0)[lb]{\smash{\SetFigFont{8}{8.0}{rm}2}}}
\put(220,300){\makebox(0,0)[lb]{\smash{\SetFigFont{12}{14.4}{it}p}}}
\put(240,300){\makebox(0,0)[lb]{\smash{\SetFigFont{12}{14.4}{rm},}}}
\put(255,300){\makebox(0,0)[lb]{\smash{\SetFigFont{12}{14.4}{rm}${\muB}$}}}
\put(500,300){\makebox(0,0)[lb]{\smash{\SetFigFont{12}{14.4}{it}p}}}
\put(520,300){\makebox(0,0)[lb]{\smash{\SetFigFont{12}{14.4}{rm},}}}
\put(540,300){\makebox(0,0)[lb]{\smash{\SetFigFont{12}{14.4}{rm}${\muC}$}}}
\put(230,290){\makebox(0,0)[lb]{\smash{\SetFigFont{8}{8.0}{rm}2}}}
\put(510,290){\makebox(0,0)[lb]{\smash{\SetFigFont{8}{8.0}{rm}3}}}
\end{picture}

\nopagebreak
\bigskip
\caption[f1]{
\label{f1}{
Box diagrams which contribute to the four photon function.
Dotted lines represent photons, and solid lines stand for electrons.}}
\end{figure}
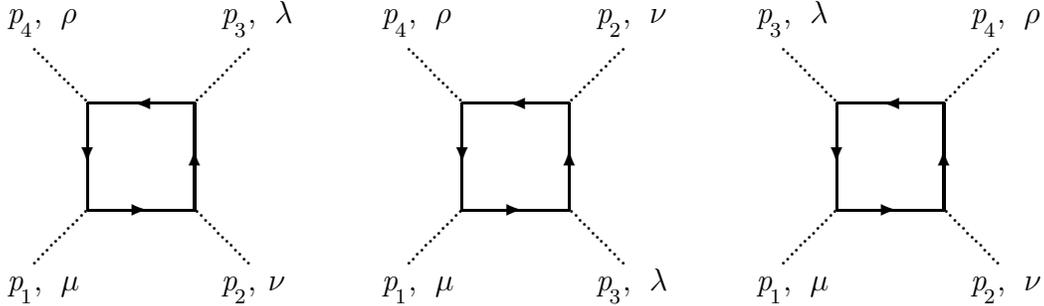

\begin{figure}[t]
\setlength{\unitlength}{0.006700in}%
\begingroup\makeatletter\ifx\SetFigFont\undefined
\def\x#1#2#3#4#5#6#7\relax{\def\x{#1#2#3#4#5#6}}%
\expandafter\x\fmtname xxxxxx\relax \def\y{splain}%
\ifx\x\y   
\gdef\SetFigFont#1#2#3{%
  \ifnum #1<17\tiny\else \ifnum #1<20\small\else
  \ifnum #1<24\normalsize\else \ifnum #1<29\large\else
  \ifnum #1<34\Large\else \ifnum #1<41\LARGE\else
     \huge\fi\fi\fi\fi\fi\fi
  \csname #3\endcsname}%
\else
\gdef\SetFigFont#1#2#3{\begingroup
  \count@#1\relax \ifnum 25<\count@\count@25\fi
  \def\x{\endgroup\@setsize\SetFigFont{#2pt}}%
  \expandafter\x
    \csname \romannumeral\the\count@ pt\expandafter\endcsname
    \csname @\romannumeral\the\count@ pt\endcsname
  \csname #3\endcsname}%
\fi
\fi\endgroup
\begin{picture}(720,204)(120,405)
%
%
\thicklines
\put(645,440){\vector( 1, 0){140}}
\put(785,440){\line( 1, 0){ 55}}
\put(120,440){\vector( 1, 0){ 65}}
\put(185,440){\vector( 1, 0){140}}
\put(325,440){\vector( 1, 0){160}}
\put(485,440){\vector( 1, 0){160}}
\multiput(240,560)(0.0,-4.1){30}{\makebox(0.9,1.3){\SetFigFont{10}{12}{rm}.}}
\multiput(400,560)(0.0,-4.1){30}{\makebox(0.9,1.3){\SetFigFont{10}{12}{rm}.}}
\multiput(560,560)(0.0,-4.1){30}{\makebox(0.9,1.3){\SetFigFont{10}{12}{rm}.}}
\multiput(720,560)(0.0,-4.1){30}{\makebox(0.9,1.3){\SetFigFont{10}{12}{rm}.}}
\put(170,410){\makebox(0,0)[lb]{\smash{\SetFigFont{12}{14.4}{it}k}}}
\put(220,580){\makebox(0,0)[lb]{\smash{\SetFigFont{12}{14.4}{it}p}}}
\put(380,580){\makebox(0,0)[lb]{\smash{\SetFigFont{12}{14.4}{it}p}}}
\put(540,580){\makebox(0,0)[lb]{\smash{\SetFigFont{12}{14.4}{it}p}}}
\put(700,580){\makebox(0,0)[lb]{\smash{\SetFigFont{12}{14.4}{it}p}}}
\put(230,570){\makebox(0,0)[lb]{\smash{\SetFigFont{8}{8.0}{rm}1}}}
\put(390,570){\makebox(0,0)[lb]{\smash{\SetFigFont{8}{8.0}{rm}2}}}
\put(550,570){\makebox(0,0)[lb]{\smash{\SetFigFont{8}{8.0}{rm}3}}}
\put(710,570){\makebox(0,0)[lb]{\smash{\SetFigFont{8}{8.0}{rm}4}}}
\put(240,580){\makebox(0,0)[lb]{\smash{\SetFigFont{12}{14.4}{rm}, ${\muA}$}}}
\put(400,580){\makebox(0,0)[lb]{\smash{\SetFigFont{12}{14.4}{rm}, ${\muB}$}}}
\put(560,580){\makebox(0,0)[lb]{\smash{\SetFigFont{12}{14.4}{rm}, ${\muC}$}}}
\put(720,580){\makebox(0,0)[lb]{\smash{\SetFigFont{12}{14.4}{rm}, ${\muD}$}}}
\put(765,410){\makebox(0,0)[lb]{\smash{\SetFigFont{12}{14.4}{it}k}}}
\put(295,410){\makebox(0,0)[lb]{\smash{\SetFigFont{12}{14.4}{it}k}}}
\put(455,410){\makebox(0,0)[lb]{\smash{\SetFigFont{12}{14.4}{it}k}}}
\put(615,410){\makebox(0,0)[lb]{\smash{\SetFigFont{12}{14.4}{it}k}}}
\put(315,410){\makebox(0,0)[lb]{\smash{\SetFigFont{12}{14.4}{rm}+}}}
\put(475,410){\makebox(0,0)[lb]{\smash{\SetFigFont{12}{14.4}{rm}+}}}
\put(635,410){\makebox(0,0)[lb]{\smash{\SetFigFont{12}{14.4}{rm}+}}}
\put(335,410){\makebox(0,0)[lb]{\smash{\SetFigFont{12}{14.4}{it}p}}}
\put(495,410){\makebox(0,0)[lb]{\smash{\SetFigFont{12}{14.4}{it}p}}}
\put(655,410){\makebox(0,0)[lb]{\smash{\SetFigFont{12}{14.4}{it}p}}}
\put(345,400){\makebox(0,0)[lb]{\smash{\SetFigFont{8}{8.0}{rm}1}}}
\put(505,400){\makebox(0,0)[lb]{\smash{\SetFigFont{8}{8.0}{rm}12}}}
\put(665,400){\makebox(0,0)[lb]{\smash{\SetFigFont{8}{8.0}{rm}123}}}
\end{picture}

\nopagebreak
\bigskip
\caption[f2]{\label{f2}{One of the four forward scattering amplitudes,
           corresponding to the first diagram in \hbox{Fig. 1}.}}
\end{figure}
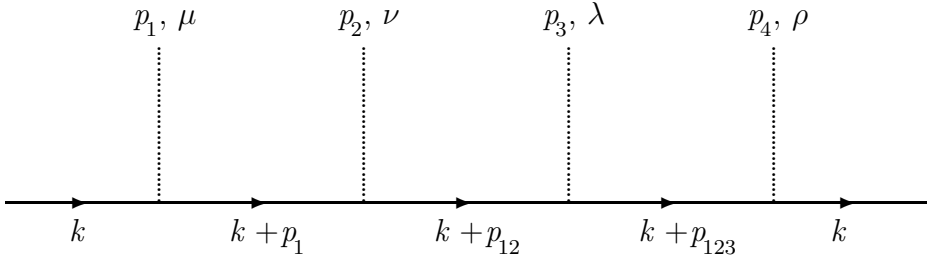

The graphs which contribute to the four photon function are shown in 
Fig. \ref{f1}. There are three other contributions obtained by
charge conjugation. To evaluate these diagrams, we use
the analytically continued imaginary-time thermal perturbation
theory \cite{kapusta:book89,lebellac:book96,das:book97}. This approach can be
formulated \cite{BFT} so as to express the thermal Greens function in
terms of {\it forward scattering} amplitudes of an on-shell fermion
in an external electromagnetic field, as depicted in Fig. \ref{f2}.
Each of these amplitudes corresponds to a
{\it cut} in one of the internal lines of the boxes in Fig. \ref{f1}.
This generates a total of {\it $4\times 6=24$ diagrams}, which can be
systematically obtained from the graph in  Fig. \ref{f2}, 
by permutations of the external momenta and
polarizations. 

The finite-temperature contribution of the box diagrams can then be
written in the form
\beN{eq2}
\Pi^{\seqi}({\seqv})=-\frac{{ e^4}}{(2\pi)^2} \int
\frac{d^2\vec k}{2\omega_k}\left(\frac 1 2 - N(\omega_k)\right)
\left\{
\sum_{ijkl}{ B}^{\mu\nu\lambda\rho}_{(ijkl)}
+ (k\leftrightarrow -k)
\right\}
\ee
Here $\omega_k=\sqrt{k^2 +m^2}$, ${N(\omega_k)}=({{\rm e}^{\omega_k/T}}+1)^{-1}$,
and the sum is over the permutations $(ijkl)$ of $(1234)$.
Each ${ B}$ has a numerator which involves a
trace over the Dirac indices.
For example
\beN{eq3}
\label{B1234}
\left.
{ B}^{\mu\nu\lambda\rho}_{(1234)}=\displaystyle{
\frac{{\rm tr}\left[
\left({{\bf \slash}\hskip-.65em\relax k+m}\right)
\gamma^{{\muA}    }
\left({{\bf \slash}\hskip-.65em\relax k+
       {\bf \slash}\hskip-.5em\relax p_{1  }+m}\right)
\gamma^{{\muB}    }
\left({{\bf \slash}\hskip-.65em\relax k+
       {\bf \slash}\hskip-.5em\relax p_{12 }+m}\right)
\gamma^{{\muC} }
\left({{\bf \slash}\hskip-.65em\relax k+
       {\bf \slash}\hskip-.5em\relax p_{123}+m}\right)
\gamma^{{\muD} }\right]}
            {\left({ 2\,k.p_{1  }   +p_{1  }^2}\right)
             \left({ 2\,k.p_{12 }   +p_{12 }^2}\right)
             \left({ 2\,k.p_{123}   +p_{123}^2}\right)}}
\right|_{k_0=\omega_k} \, ,
\ee
where $p_{12} =p_1 +p_2$, etc. 
Here, we are only interested in the contributions from the trace
in Eq. (\ref{B1234}) which contain {\it odd powers of the mass}, since these
will lead to parity-breaking terms.

Let us study first the zero temperature contribution, which is
associated with the factor $1/2$ in the first bracket of
Eq. (\ref{eq2}), as $N(\omega_k)$ vanishes in this limit. The
computation can be performed explicitly in the {\it low momentum}
region, where $|p_\mu|\ll m$. The result can then be expressed in 
terms of a series in powers of $p/m$, which begins with the leading 
contribution
\beN{eq4}
\bmt{lll}
\Pi^{\seqi}_{T=0} & = & -\displaystyle{\frac{{i\,e^4}}{16\pi\,m^6}} 
\left[
\ep{\muA}{\muB}{\alpha}p_1^\alpha\,(p_2)^2 + 
\ep{\muA}{\alpha}{\beta}p_1^\alpha\,p_2^\beta\,p_2^{\muB}
\right] \\ \\
& \times & 
\left[
\eta^{\muC\muD}p_3\cdot p_4 - p_3^{\muD} p_4^{\muC}
\right] + {\rm permutations}
\emt
\ee
It is interesting to note that this result is consistent with the
Coleman-Hill theorem \cite{coleman:1985zi} which implies that, in the
four point Greens function, at zero temperature, the terms of order $p$
should be absent. In fact, the above structure shows that the
parity-violating 
contributions, generated by the box at $T=0$, begin only with terms of
order $(p/m)^5$. In the configuration space, the low-energy
effective action associated with Eq. (\ref{eq4}) can be written in the form
\beN{eq5}
\Gamma^4_{T=0} =
-\frac{{e^4}}{64\pi\,m^6}\intn{3}\ep{\mu}{\nu}{\lambda}
F_{\mu\nu}\left(\partial^\tau F_{\tau\lambda}\right)
F^{\rho\sigma} F_{\rho\sigma},
\ee
which is manifestly Lorentz and gauge invariant ({\it small} and {\it
large}).

We now turn to the parity-violating, finite temperature contributions
generated by the box diagram. Such contributions are
non-analytic \cite{das:book97}, so that the thermal behavior is
quite  distinct in different regions under consideration. For example, in the 
{\it long wavelength} limit,
$\vec p_i =0$, these contributions can be expressed as
\beN{eq6}
\Pi^{\seqi}_{LW} = i\,e^4
\left(\eta^{\muA\muB}-\delta^\muA_0\delta^\muB_0\right)
\ep{0}{\muC}{\muD}\left(p_3^0-p_4^0\right)\, R_{34}\left(p_i^0,T\right) +
{\rm permutations},
\ee
where the functions $R_{mn}$ have, in general, a rather complicated
structure depending upon external energies and the temperature.

A great simplification occurs in the low-energy region where 
$p^0_i\ll m$, when those functions reduce to a common form given by
\beN{eq7}
R =\frac{m\,T}{64}\,\sum_{l=-\infty}^{\infty}
\left\{
\left(\frac{5\,m^2}{\Delta_l^6}+\frac{3}{\Delta_l^4}\right)
\ln{\left(1+\frac{\Delta_l^2}{m^2}\right)}
-\frac{5}{\Delta_l^4} -\frac{1}{2\,m^2\Delta_l^2}
\right\},
\ee
where $\Delta_l \equiv (2l+1)\,\pi\,T$. In the high temperature
limit, the leading contribution comes from the last term in
Eq. (\ref{eq7}). Performing the summation over $l$, we then obtain
that
\beN{eq8}
R(T\gg m) = \frac{1}{512}\frac{1}{m\,T}.
\ee
It is worth remarking here that the only non-vanishing components of
the amplitude in Eq. (\ref{eq6}), are the ones when the indices take
spatial values. Using this fact, we see that the leading
contribution comes from an extensive effective action of the form
\beN{eq9}
\tilde\Gamma^4_{LW} = \frac{e^4}{512\,m\,T}
\intn{3}\epsilon_{0ij}
E_i\left(\partial_t^{-1}E_j\right)
\left(\partial_t^{-1}E_k\right)
\left(\partial_t^{-1}E_k\right),
\ee
where $\vec E$ denotes the electric field. This action is non-local and
manifestly gauge invariant.

Next, we  discuss the thermal behavior of the box in the static
limit where $p_{i}^{0}=0$. In this case, due to the very complicated
angular  integrations,
the calculations are extremely difficult, even when using 
{\it computer algebra}. As a result, we
have restricted ourselves to a special 3-momentum configuration of the external
momenta, where $\vec p_1=\vec p_2=\vec p_3=\vec p$ (because of
momentum conservation, $\vec p_4=-3\vec p$). We then find that the
only non-vanishing components of the amplitude are
\beN{eq10}
\Pi^{000i_4}_S=\frac{1}{4\,|\vp|^2} p_i\epsilon_{0\;i\;i_4} 
\Pi_1(\vp,T)
\ee
and, 
\beN{eq11}
\Pi^{0 i_2 i_3 i_4}_S=\frac{1}{12\,|\vp|^4} p_i\epsilon_{0\;i\;i_4}
\left(p_{i_2} p_{i_3} - |\vec p|^2 \delta_{i_2 i_3}\right)
\Pi_2(\vp,T),
\ee
where $\Pi_{1,2}(\vp,T)$ are rather complicated functions of the momenta
and the temperature. However, an important simplification occurs in
the low momentum region $|\vp|\ll m,T$. In this domain, 
$\Pi_2 = O(p^6/m^6)$ becomes negligible, while the expression of
$\Pi_1$ reduces to
\beN{eq12}
\tilde \Pi_1(\vp,T) =
\frac{6\,i\,e^4}{4 \pi}\left[\thm - \thmN{3}\right]\frac{|\vp|^2}{T^2}
+ O\left(\frac{|\vp|^4}{m^2\,T^2}\right)
\ee
In the high temperature limit, the leading term is of order $1/T^3$,
which is quite different from the $1/T$ behavior of 
the result in the {\it long wavelength} limit  (\ref{eq8}).

The leading contribution given in Eqs. (\ref{eq10}) and (\ref{eq12})
can be associated with the effective non-extensive action
\beN{eq13}
\tilde\Gamma^4_S=\frac{e^4}{4 \pi\,T^2}
\left[\thm - \thmN{3}\right]\intn{3}\,a^3\,B,
\ee
where 
\[
a = \int_0^{1/T}{\rm d}t\, A_0(t,\vec x)
\]
and $B=B(\vec x)$ is the static magnetic field.
This form is consistent with the result derived from the all-orders
effective action, which was obtained earlier in the special gauge background.

In contrast to the effective action obtained in the long wavelength limit 
[Eq. (\ref{eq9})], the static action (\ref{eq13}) 
{\it is not invariant} under 
{\it large gauge transformations} generated by 
$a\rightarrow a + 2\pi\,N$, where $N$ is an integer winding
number. (Incidentally, at finite temperature, the effective action is
non-unique \cite{DH} and depends on the limits in which it is derived.) But
one can derive, in this case, a Ward identity for {\it large gauge}
invariance, which relates the amplitudes obtained in perturbation
theory. Motivated by the structure of Eq. (\ref{eq13}), let us write
the all-orders static effective action in the form
\beN{eq14}
\tilde\Gamma_S=\frac{e\,T}{2 \pi}\intn{3}\,\tilde\Gamma(\tilde a)B,
\ee
where $\tilde a = e a$. It has been noted in \cite{DD} that in
the special background $A_0=A_0(t)$ and $\vec A=\vec A(\vec x)$,
$\tilde\Gamma(\tilde a)$ corresponds precisely to the {\it real part}
of the effective action $\Gamma^{(1)}(\tilde a)$ which describes the
behavior of the $0+1$ dimensional theory.
This action obeys, for a single fermion flavor, the {\it large gauge
  Ward identity} \cite{das:1999rc}
\beN{eq15}
\frac{\partial^2\Gamma^{(1)}}{\partial\tilde a^2}=
i\left[\frac 1 4 - 
\left(\frac{\partial\Gamma^{(1)}}{\partial\tilde a}\right)^2\right],
\ee
where the one point function has the value
\beN{eq16}
\left.\frac{\partial\Gamma^{(1)}}{\partial\tilde a}
\right|_{\tilde a = 0} =
\frac 1 2 \thm.
\ee
Using these relations, as well as the conditions following from them, 
we find after some analysis that 
$\tilde\Gamma(\tilde a)=\Re\left[\Gamma^{(1)}(\tilde a)\right]$
satisfies the large gauge Ward identity
\beN{eq17}
\frac{\partial^2\tilde \Gamma}{\partial\tilde a^2}=
\frac{1}{{\rm sinh}(m/T)} 
\frac{\partial\tilde \Gamma}{\partial\tilde a}\,
\sin{\left(2\tilde\Gamma\right)},
\ee
which reflects the large gauge invariance of the static ${\rm QED_3}$
theory, in the leading approximation. The nonlinear relation
(\ref{eq17}), which can be checked perturbatively, shows that all
amplitudes are related recursively to the one-point function. The
solution of Eq. (\ref{eq17}) is given by
\beN{eq18}
\tilde\Gamma(\tilde a)=
{\rm arctan}\left[\thm\tan\left(\frac{\tilde a}{2}\right)\right].
\ee
Substituting this form in the expression (\ref{eq14}), we obtain 
a result which agrees, in the static limit of ${\rm QED_3}$, with the
one proposed earlier for the parity-breaking effective action.

This work was supported in part by U.S. Dept. Energy Grant DE-FG
02-91ER40685, NSF-INT-9602559 as well as by CNPq, Brazil.


\begin{thebibliography}{10}
\bibitem{dunne:1988lh} A nice review of the subject with many references
can be found in, G. Dunne, {\it Topological Aspects of Low Dimensional
Systems}, Les Houches Summer School, 1998.
\bibitem{babu:1987rs}
K.~S. Babu, A. Das and P. Panigrahi, Phys. Rev. {\bf D36},  3725
(1987); E. Poppitz, Phys. Lett. {\bf B252}, 417 (1990); 
I. J. R. Aitchison, C. Fosco and J. Zuk, Phys. Rev. 
{\bf D48}, 5895 (1993).
\bibitem{deser:1982} S. Deser, R. Jackiw and S. Templeton,
Ann. Phys. {\bf 140},372 (1982).
\bibitem{DDL} G. Dunne, K. Lee and C. Lu, Phys. Rev. Lett. {\bf 78}, 3434
(1997). 
\bibitem{DD}
A. Das and G. Dunne, Phys. Rev. {\bf D57},  5023  (1998).
\bibitem{fosco:1997ei}
S. Deser, L. Griguolo and D. Seminara, Phys. Rev. Lett. {\bf 79}, 1976
(1997);
C. Fosco, G.~L. Rossini and F.~A. Schaposnik, Phys. Rev. Lett. {\bf 79},  1980
   (1997)
and Phys. Rev. {\bf D56},  6547 (1997);
I. J. R. Aitchison and C. D. Fosco, Phys. Rev. {\bf D57}, 1171 (1998)
\bibitem{DA} A. Das and A. J. da Silva, Phys. Rev. {\bf D59}, 105011 (1999).
\bibitem{BDF} F. T. Brandt, A. Das and J. Frenkel, in preparation.
\bibitem{kapusta:book89}
J.~I. Kapusta, {\em Finite Temperature Field Theory} (Cambridge University
  Press, Cambridge, England, 1989).
\bibitem{lebellac:book96}
M.~L. Bellac, {\em Thermal Field Theory} (Cambridge University Press,
  Cambridge, England, 1996).
\bibitem{das:book97}
A. Das, {\em Finite Temperature Field Theory} (World Scientific, NY, 1997).
\bibitem{BFT}
J. Frenkel and J.~C. Taylor, Nucl. Phys. {\bf B374},  156  (1992); \\
F.~T. Brandt and J. Frenkel, Phys. Rev. {\bf D56},  2453  (1997).
\bibitem{coleman:1985zi}
S. Coleman and B. Hill, Phys. Lett. {\bf B159},  184  (1985).
\bibitem{DH} A. Das and M. Hott, Phys. Rev. {\bf D50}, 6655 (1994).
\bibitem{das:1999rc}
A. Das, G. Dunne and J. Frenkel, Phys. Lett. {\bf B472},  332  (2000).
\end{thebibliography}
\end{document}